\documentclass{jfm}

\usepackage{graphicx}
\usepackage{natbib}
\usepackage{amssymb}
\usepackage{amsmath,amsfonts}

\usepackage{float}
\usepackage{ulem}
 \usepackage{color}

\ifCUPmtlplainloaded \else
  \checkfont{eurm10}
  \iffontfound
    \IfFileExists{upmath.sty}
      {\typeout{^^JFound AMS Euler Roman fonts on the system,
                   using the 'upmath' package.^^J}%
       \usepackage{upmath}}
      {\typeout{^^JFound AMS Euler Roman fonts on the system, but you
                   dont seem to have the}%
       \typeout{'upmath' package installed. JFM.cls can take advantage
                 of these fonts,^^Jif you use 'upmath' package.^^J}%
      }
  \else
  \fi
\fi

\ifCUPmtlplainloaded \else
  \checkfont{msam10}
  \iffontfound
    \IfFileExists{amssymb.sty}
      {\typeout{^^JFound AMS Symbol fonts on the system, using the
                'amssymb' package.^^J}%
       \usepackage{amssymb}%
         
         \let\geq=\geqslant
      }{}
  \fi
\fi

\ifCUPmtlplainloaded \else
  \IfFileExists{amsbsy.sty}
    {\typeout{^^JFound the 'amsbsy' package on the system, using it.^^J}%
     \usepackage{amsbsy}}
    {\providecommand\boldsymbol[1]{\mbox{\boldmath $##1$}}}
\fi

\newsavebox{\astrutbox}
\sbox{\astrutbox}{\rule[-5pt]{0pt}{20pt}}

\usepackage{color}

\title[Unsteady wave pattern generation by water striders]{Unsteady wave pattern generation by\\ water striders}

\author[T. Steinmann, M. Arutkin, P. Cochard, E. Rapha\"el, J. Casas and M. Benzaquen]%
{Thomas Steinmann$^{1}$\thanks{E-mail address first author: thomas.steinmann@univ-tours.fr \quad\quad\quad\quad\quad \ \, }, 
  Maxence Arutkin$^2$,  {Pr\'ecillia Cochard$^{1,3}$}, Elie Rapha\"el$^2$, J\'er\^ome Casas$^{1}$ and Michael Benzaquen$^4$\thanks{E-mail address for correspondence: michael.benzaquen@polytechnique.edu}  }

\affiliation{$^1$Institut de Recherche sur la Biologie de l'Insecte, UMR CNRS 7261, Universit\'e de Tours, France\\[\affilskip]
$^2$UMR CNRS 7083 Gulliver, ESPCI Paris, 10 rue Vauquelin, 75005 Paris, France\\[\affilskip]
{$^3$Current adress: D\'epartement de biologie, 
Universit\'e Laval, Qu\'ebec,
QC G1V 0A6, Canada}\\[\affilskip]
$^4$LadHyX, UMR CNRS 7646, 
Ecole  Polytechnique, 91128 Palaiseau Cedex, France}

\pubyear{2010}
\volume{650}
\pagerange{119--126}
\begin{document}
\maketitle

\bigskip

\begin{abstract}
We perform an experimental and theoretical study of the wave pattern generated by the leg strokes of water striders during a propulsion cycle. Using the synthetic Schlieren method, we are able to  measure the dynamic response of the free surface accurately. In order to match experimental conditions, we extend B\"uhler's  theory of impulsive forcing (\cite{buhler}) to finite depth. We demonstrate the improved ability of this approach to reproduce the experimental findings, once the observed continuous forcing and hence non-zero temporal and spatial extent of the leg strokes is also taken into account.
\end{abstract}

\bigskip\bigskip\bigskip\bigskip

\tableofcontents

\newpage

\section*{Introduction}

Life at the air-water interface requires specific adaptations, as displayed by several insect and arachnid species living at this interface and exploiting the static and dynamic deformations of water (\cite{suter2013}). Water striders, or \textit{gerrids}, are among the best known insects operating at the air-water interface.
 The locomotion of these insects in this habitat has been extensively studied (\cite{denny2004paradox,Bush2006,vogel2013}), and  the discovery that they can jump into the air, using the water surface as trampoline (\cite{Koh2015,ortega2017}), has excited additional considerable interest. Water striders are also increasingly being used as a template for microrobot design (\cite{Hu2007,Yuan2012,Yang2016b,Yang2016a,song2007surface}).\medskip

{Locomotion at the air-water interface reveals a number of physical phenomena}.
{In particular, the displacement of a disturbance at the surface creates a complex wave pattern (\cite{Raphael1996}). The waves and vortices in the strider's wake {are} the signature of the momentum applied to the water. Other insects  produce waves upstream  from their bodies (\cite{Voise2010, xu2012experimental, jia2015energy}). 
Whirligig beetles, or \textit{gyrinids} produce upstream waves and are thought to use  capillary waves to explore their environment through echolocation (\cite{voise2014echolocation,tucker1969wave}). Interfacial 2D flight is a recently discovered phenomenon in which  beetles fly over water while keeping the tips of their hindlimbs attached to the water surface to increase stability (\cite{mukundarajan2016surface}).
  Theoretical studies of the formation of waves and vortices at the interface have advanced our knowledge considerably, particularly for water striders (\cite{hu2003hydrodynamics, buhler,denny2004paradox,gao2011numerical,zheng2015modeling}). {However}, it remains unclear whether surface waves are more important than vortices for transferring the momentum imparted by the leg to the water. 
Advances in experimental research have been based principally on the work of \cite{hu2003hydrodynamics, hu2010hydrodynamics} and \cite{Rinoshika2011}, who performed PIV (particle imaging velocimetry) measurements on the water surface, {but only from below} (see also \cite{zheng2015modeling}). They also filmed the animal from the side, thereby obtaining  a rough estimate of the dip of the meniscus. However, none of these studies recorded the surface wave signature in detail or involved side-on PIV recordings. The calculations of \cite{gao2011numerical} predict the detachment of the vortices  from the water surface under certain conditions, whereas the experimental work of \cite{Rinoshika2011} showed that the vortices did not detach. Unfortunately, these experiments were carried out in very shallow water (1.2 mm deep, with a 0.3 mm layer of ink at the bottom), making it difficult to reconcile the results of these experiments with the assumption of infinite depth applied in the available theoretical studies. The detachment or non-detachment of vortices from the surface has major implications for estimation of the thrust generated. Clear answers to this question are therefore required in the context of a force balance. Experimental quantification of {\it i)} the kinematics of  water strider legs, {\it ii)} the surface deformation, and {\it iii)} vortex production are therefore {required}, together with {\it iv)} an extension of the theory to shallow water, given that most experiments are performed in such conditions. \medskip

This paper constitutes a key advance in this domain: we quantify, with unmatched precision (5\,$\mu m$ resolution), the deformations occurring during {the} {leg strike} of the water strider, to improve our understanding of the importance of accounting for the details of the fluid-structure interaction, as opposed to the impulsive forcing model of B\"uhler, in which the strike is localised in time and space. We also extend  B\"uhler's theory (\cite{buhler}) to shallow water and continuous non-impulsive forcing. Our work has implications beyond  locomotion at the air-water interface, as small-amplitude wave-trains are used by many insects living at the water surface, for sexual communication and predation \cite{wilcox1972communication,wiese1974mechanoreceptive,bleckmann1994stimulus}. Furthermore,  B\"uhler's theory is increasingly being applied in other contexts, such as impacts on films \cite{basu2017angled} or the energy contained in the breaking of waves \cite{pizzo2016current}. Our theoretical results are, therefore, likely to be applicable beyond the case study analysed here.

\section{Experiments}
\subsection{Insect collection}

This study was conducted on the water strider \textit{Gerris paludum}. We collected adult and larval stages regularly from April to August 2014, from a freshwater pond in Tours, France 
($47^{\circ}22'01.3''\text{N}  \,-\, 0^{\circ}41'29.5''\text{E}$). 
The insects were kept in aquariums (37$\times$17\,cm) at a constant temperature of $23\pm 2^{\circ}$C, under a natural light cycle, until their use in experiments. We provided frozen fruit flies for feeding every other day. We validated and calibrated the profilometry technique on 70 individuals selected for \textit{static} measurements. We established five groups for immature stages (1st to 5th {instar}\footnote{{The word \textit{instar} designates an insect in any one of its periods of postembryonic growth between moults.}}) and one group of adults, based on body length. Large adults generated large waves, with large surface gradients, during their leg strokes. For  \textit{dynamic} measurements, the analysis was restricted to two groups of instars (8 individuals in total from the two groups: six 2nd instars and two  3rd instars), because the synthetic Schlieren method (\cite{moisy2009synthetic}) (detailed below) is accurate only for small surface-profile slopes. 
The waves generated by first-instar insects were undetectable. {We believe, but cannot definitively demonstrate, that the absence of visible wave generation during first-instar propulsion is the consequence of the limited resolution of our measurement technique (see below)\footnote{{Note that early-instar striders move their legs at a speed below the critical velocity of 23\,cm/s (see \cite{Raphael1996}), falling into the framework of DennyÕs paradox on the impossibility of wave generation.
Note, however, that this situation applies only to straight and steady motion. Here, we show experimentally that strider leg motion is neither regular nor rectilinear, and that there should, therefore, always be waves, even if their amplitude is too small for detection. In addition, \cite{hu2010hydrodynamics} argued that the smallest water walkers, generating negligible deformations of the surface, make use principally of microscale contact forces generated by the brushing of the legs against the air-water interface. The ratio between leg curvature and the brushing forces is dependent on the morphological and functional characteristics of the tarsi and their integuments (\cite{sun2018study}).
}}}.

\subsection{Synthetic Schlieren method}
 
Several profilometry techniques have been developed for measuring the local topography of the air-water interface. We chose to use the synthetic Schlieren method initially developed by \cite{moisy2009synthetic}. This innovative method is based on the capture, from above, of high-speed video camera  images of a random dot pattern at the bottom of the basin transmitted through the liquid interface. Changes in the topography of the liquid-air interface result in changes in the refracted dot pattern images. The insect is also captured by the camera, albeit in blurred images, because the camera has a limited depth of field and is is focused on the dot pattern at the bottom of the basin. Below,  \textit{$\zeta $}(\textit{x, y}) denotes the free surface displacement with respect to the equilibrium position, and  \textit{$\delta $}\textbf{\textit{r}}(\textit{x, y}) denotes the resulting change in refracted dot pattern. \cite{moisy2009synthetic} demonstrated that, for small surface gradients, namely {$\nabla \zeta \ll 1$, or equivalently $\zeta \ll \lambda$, where $\lambda$ denotes the smallest wavelength}, the distortion field \textit{$\delta $}\textbf{\textit{r}} is simply proportional to the free surface gradient. As in particle image velocimetry (PIV) techniques, digital image correlation algorithms can then be used to determine the apparent displacement field between the refracted image and the reference image obtained when the surface is flat. Finally, numerical integration of this displacement field, based on a least-squares inversion of the gradient operator, can be used to reconstruct the instantaneous surface height. 
This technique and the mathematical simplification used during reconstruction of the surface height are valid only for the {first-order paraxial angle} approximation  (see \cite{moisy2009synthetic}), implying that the camera must be placed at a large distance from the target.

\subsection{Static measurements}

As described by \cite{moisy2009synthetic}, we first validated the full measurement procedure, using a solid Plexiglas model of a wavy surface of wavelength $20 \,$mm and amplitude $0.5 \,$mm, which we borrowed from F. Moisy. We carried out the \textit{static} experiments in a $185\times125\,$mm plastic tank, filled with tap water ($9 \text{\,mm}$ deep for adults and $3 \text{\,mm}$ deep for instars).
Images of the water surface were taken with a $1632 \times1200$ pixels high-speed camera (Vision Research Phantom V9.0) located above the tank. The distance $H$ between the water surface and the high-speed camera,  and the resulting width $L$ of the field of view  were adapted according to insect body length. We used $H_\text{ad} = 71 \text{\,cm}$ and $L_\text{ad} = 27.5 \text{\,mm}$ for adults,  $H_\text{la.5,4} = 45 \text{ cm}$ and $L_\text{la.5,4} = 15 \text{ mm}$  for 5th  and 4th instars,  $H_{\text{la.3}} = 39 \text{ cm}$ and $L_{\text{la.3}} = 12 \text{ mm}$ for 3rd instars, and  $H_\text{la1,2} = 36 \text{ cm}$ and $L_\text{la1,2} = 7 \text{\,mm}$ for 1st and 2nd instars. 
For each experiment, an insect was placed in the tank and recorded when it was within the field of view of the camera and resting at the surface with no water movement. We estimated the displacement field \textit{$\delta $}\textbf{\textit{r}} of the dot pattern, using a  digital image correlation algorithm, with interrogation windows of 16 \textit{$\times$}16 pixels for the calculation of correlation functions. With such small interrogation windows, and the use of classical subpixel interpolation, it was possible to achieve a resolution of 0.1 pixels for the estimation of dot pattern displacement (\cite{feldmann2001optical}). We thus obtained a precision for the estimation of \textit{$\zeta $}(\textit{x, y}) of \textit{$\pm$}6.2 $ \mu m $ for adults and \textit{$\pm$}1.2 $ \mu m $ for young instars.\\
%

\begin{figure}
\centering
\includegraphics[scale=0.55]{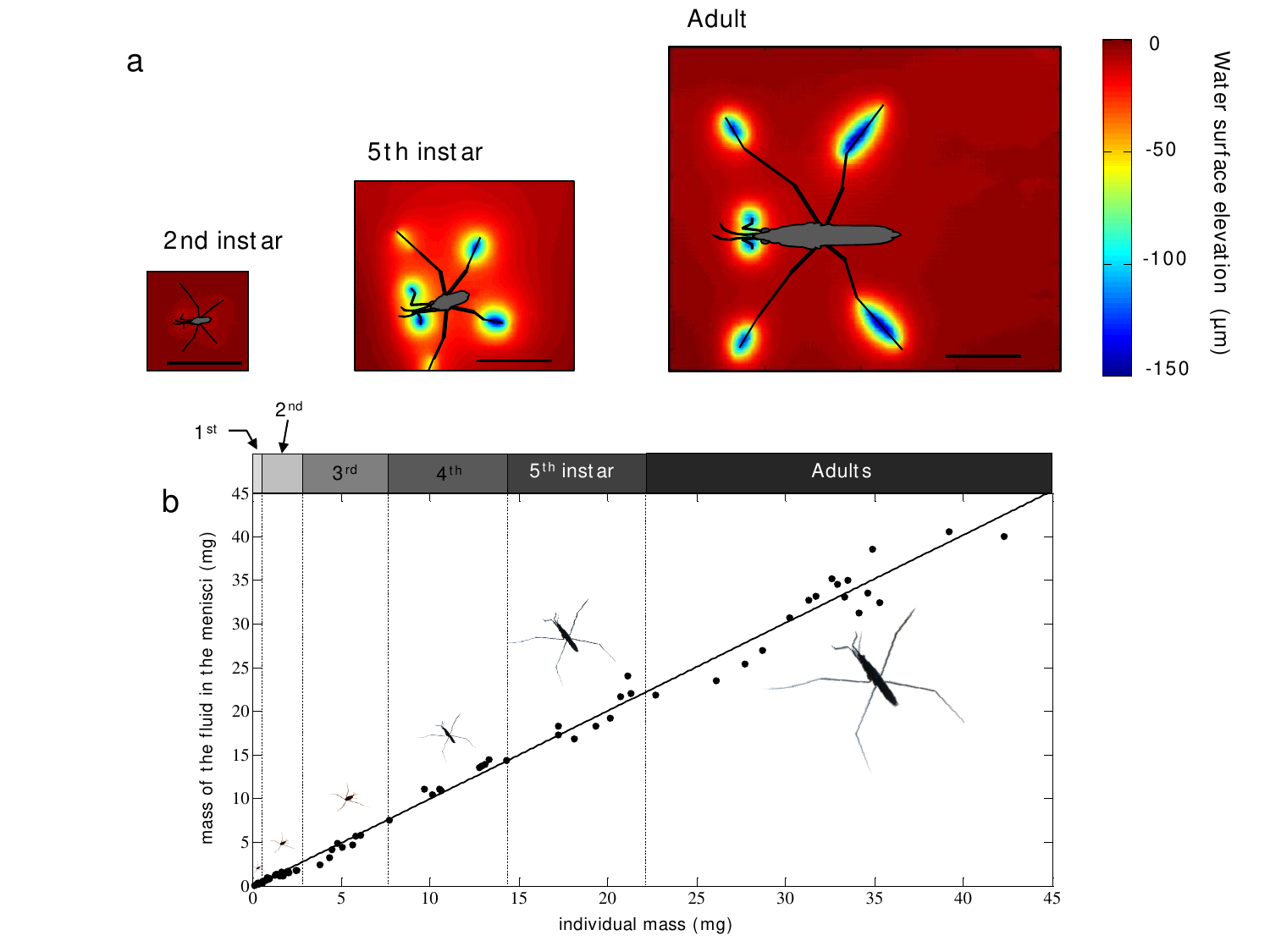}
\caption{a. Static water surface topography deformation around three different instars of non-moving \textit{Gerris paludum} . The colour scale indicates the elevation of the water surface. The scale is indicated by the black line (5 mm). b. {Mass of  displaced fluid  in the menisci as a function of insect mass, from first instar to adult.} Linear regression results: ${t }= 1{.}0401$, ${df }= 68$, ${p\_}\text{value} = 0\textit{.}30$. 
 }
\label{F2}
\end{figure}

According to the vertical force balance of a water strider in a static situation  (see \cite{hu2010hydrodynamics} and \cite{keller1998surface}), the mass of the volume displaced by the insect at the free surface is equal to the mass of the insect. Indeed, given the free surface displacement $\zeta(x,y)$ resulting from an external pressure field $p_\text{ext}(x,y)$, with $\hat\zeta(k)$ and $\hat p(k)$ as the corresponding Fourier transforms, the volume of the displaced fluid $V_\text{d}$ can be linked directly to $\zeta(x,y)$ as follows:
\begin{eqnarray}
V_\text{d}=\iint \text{d}x\text{d}y \text{ } \zeta(x,y)= -\underset{k \rightarrow 0}{\text{lim}} \hat\zeta(k) \ . \label{Archimede1}
\end{eqnarray}
\noindent In addition, as $\hat\zeta(k)$ is proportional to $\hat p(k)$ (see {\it e.g.} \cite{Raphael1996}), and  $\hat p(0)=M g$, {it can be shown that}:
\begin{eqnarray}
V_{d}=\frac{M}{\rho} \ , \label{Archimede2}
\end{eqnarray}
where $M$ is the mass of the insect and $\rho $ denotes the water density. Interestingly, Archimedes law is not modified by capillarity  (see Eq.~(\ref{Archimede2})).
Individual insects were weighed separately, as follows. The insects were frozen, defrosted and weighed, immediately after defrosting, with a microbalance (Sartorius Supermicro S4). They were placed on absorbent paper during defrosting, to remove any excess water. The weight of 70 individuals (from first instars to adults), as measured with the microbalance, was compared with that predicted by the Schlieren technique. Excellent agreement was achieved, for all instars (see Fig.~\ref{F2}).
The shade thrown over the measurement field by the body of the insect might have been expected to be problematic for the evaluation of surface elevation. {Fortunately, the deformation of the pattern does not seem to be significantly  blurred by the presence of the body.}  The Schlieren technique can, therefore, be considered  an accurate and non-intrusive technique appropriate for this task (for an alternative optical method based on shadows, see \cite{zheng2016elegant}).

\subsection{{Dynamic} measurements}

\begin{figure}
\centering
\includegraphics[scale=0.39]{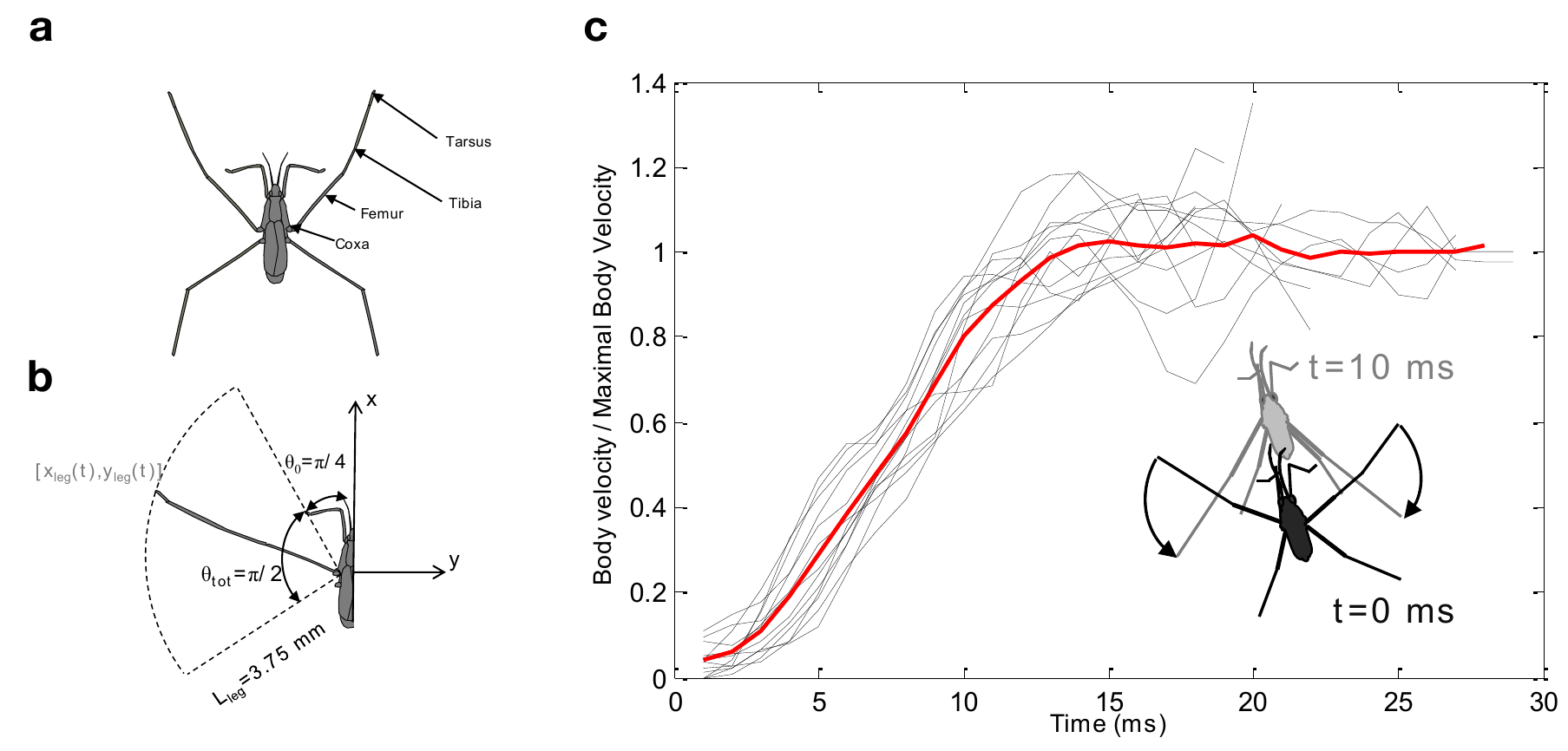}
\caption{Body and leg kinetics during a stroke.  a. Definition of the morphology of an insect leg. b. The water strider body is oriented along the $x$ axis. At rest, the 3.75 mm-long median leg makes a $\pi/4 $ angle with the $x$ axis. During the 10 ms stroke, the leg rotates by approximately $\pi/4 $ rad.  c. Dimensionless velocity of the body for 10 strokes. The small superimposed sketch superimposed illustrates the movement of an individual, over a period of 10 ms.  
}
\label{F4}
\end{figure}

 We filmed individuals propelling themselves in the middle of a shallow tank ($7\times 5$ cm). We acquired images with a 1632 $\times$ 1200 camera, and divided the full image into 101 $\times$ 74 interrogation areas of 16 $\times$ 16 pixels. The distance between  interrogation areas determines the spatial resolution of the elevation surface, {\it i.e.}  the distance between estimates of elevation  $\delta d_{s}$ ($\mu m$). The spatial resolution of the surface elevation $\delta d_{s}$  can be adjusted modifying the distance between the camera and the water, and also depends on  {\it i)} the focal distance of the optical system, {\it ii)} the distance between {the} lens and CMOS sensor and the camera and {\it iii)} the size and resolution of the CMOS sensor. Table~\ref{tab} (see Appendix B) reports the mass of the individuals, their velocities and their accelerations. In this table, we also provide the distance between the camera and the water surface, $h_\text{camera}$ (m), the resting water surface {elevation} $h_\text{water}$, the spatial resolution $\delta d_{s}$ of the images ($\mu m/\text{pixel}$), the spatial resolution of {the} elevation \textit{$\delta d_s$} ($\mu m$) and the resulting resolution of the elevation  $\eta_\text{min} = \delta d_{s} \nabla \zeta_\text{min}$,
where $\nabla \zeta_\text{min}$ denotes the minimal surface gradient that can be measured.  In turn,  $\eta_\text{min}$ is dependent on the minimum measurable random dot pattern displacement  {$\delta x_\text{min}$} and  {the} distance $h^*$, where: 
\begin{equation} \label{GrindEQ__2_3_} 
h^*=\left(\frac{1}{\alpha h_\text{water}}+\frac{1}{h_\text{camera}}\right)^{-1} ,
\end{equation} 
and  $\alpha $ is the refractive index of water.
{Using a two-dimensional FFT correlation algorithm providing subpixel interpolation,}  we measured surface displacements with a resolution of 0.1 pixels (\cite{feldmann2001optical}), the real size of a pixel being given by the spatial resolution \textit{$\delta $x}, such that $\delta x_\text{min}=0.1\delta x$, and:
\begin{equation} \label{GrindEQ__2_2_} 
\nabla \zeta_\text{min}=\frac{0.1\delta x}{h^*} \ .
\end{equation} 
 By combining equations \eqref{GrindEQ__2_3_} and \eqref{GrindEQ__2_2_}, we can see that increasing the water depth at rest $h_\text{water}$ results in a decrease in the minimum measurable slope $\nabla \zeta_\text{min}$ and, thus, a decrease in elevation resolution  $\eta_\text{min}$. {Furthermore,  as ${h}_\text{water}$ increases, the refracted image of the pattern above the wave crests includes increasingly elongated patches, particularly for large waves (\cite{moisy2009synthetic}). These patches are very difficult to analyse with a cross-correlation algorithm. We expected to see relatively large waves for large 3${}^{rd}$ instar individuals, resulting in large surface gradients, and we therefore had to keep $h_\text{water}$ relatively low (1 to 5 mm) while increasing the distance between the camera and the surface $h_\text{camera}$ , to increase $\delta d_{s}$. The balance between the spatial resolution $\delta d_{s}$ and the minimal measurable slope  $\nabla \zeta_\text{min}$ resulted in an almost constant resolution of elevation  $\eta_\text{min}$ for all measurements. \\
 
 We recorded 10 single strokes made by eight different individuals with masses of 0\textit{.}127 mg to 4\textit{.}85 mg, at 1000 frames per second, and fixed the shutter time to 500 $\mu s$  to ensure that the position of the fast-moving middle legs was accurately estimated. We determined position at various time points of the body and the two middle legs,  the joint between the body and the coxa,  the joint between the femur and tibia and the end point of the tarsus.  Figure \ref{F4}(c) {shows} the dimensionless velocity of the body during the 10 strokes, the velocity being normalised by the final constant velocity of the body. The averaged dimensionless velocity (red curve) shows that the typical duration of a stroke $\tau_\text{imp} $ ranged between 10 and 15 ms. By performing several kinetic measurements, we were able to evaluate data dispersion, which we found to be limited. For the sake of clarity, we then chose the most representative measurement for comparison with the theory (see Sect.~\ref{theorysec}).

\section{Surface response to an impulsive pressure field}
\label{theorysec}

In this section, we adapt B\"uhler's calculation for the fluid interface response to an impulsive leg stroke (given by a delta function in time) of the water strider to the experimental conditions, by extending it to finite depth. In particular, we provide an analytical expression for the surface displacement  and generalise B\"uhler's results to finite depth, to take into account typical experimental conditions. We also extend B\"uhler's impulsive approach to a rapid leg stroke (finite time $\tau_\text{imp}$). We denote $\boldsymbol r =(x,y,z)$ and $\boldsymbol x =(x,y)$.

\subsection{Infinite depth}

 Consider an incompressible, inviscid, infinitely deep liquid with a free surface at rest located at $z=0$. As stated above, B\"uhler's original model is based on the assumption that the stroke can be considered impulsive and localised. We shall also assume that the impulsive force is horizontal, aligned with the $x$ axis.
The impulsive force field applied to the fluid is taken to be of the form:
 \begin{eqnarray}
\boldsymbol{F}(\boldsymbol{r},t)&=&\boldsymbol{f}(\boldsymbol{r})\delta(t) \ , \label{expF}
\end{eqnarray}
where the spatial part $\boldsymbol f$ can be expressed as follows:
 \begin{eqnarray}
\boldsymbol{f}(\boldsymbol{r})=f_0 \delta(x)\delta(y)\delta(z+h) \boldsymbol e_x\ , \label{expf}
\end{eqnarray}
and where $h$ typically corresponds to the depth of the meniscus created by the water strider at rest. It should be noted, however, that, in this model, the free surface is considered to be flat at $t=0$, ignoring the static deformation of the interface\footnote{{This simplifying assumption is discussed in Section 4.1. }}. The resulting pressure field $p$ can be calculated from the Poisson equation:  
 \begin{eqnarray}
\Delta p&=&\boldsymbol \nabla \cdot  \boldsymbol f \ ,  \label{Poisson}
\end{eqnarray}
subject to the boundary conditions $p(x,y,z=0)=0$ and $\boldsymbol \nabla p|_{r\rightarrow \infty}\rightarrow \boldsymbol 0$, where $r = |\boldsymbol r|$.
Equation~\eqref{Poisson} takes into account the divergence of the linearised Euler equation and the condition of incompressibility. Green's function of Eq.~\eqref{Poisson} can be calculated by the method of images to ensure a vanishing solution at the interface  {(see  \cite{buhler})}. By convolving the resulting Green's function with the right-hand side of Eq.~\eqref{Poisson}, we obtain the impulsive pressure field $p$, which, at first order in $h$ can be expressed as follows:
\begin{eqnarray}
p(\boldsymbol r) &=& -h\dfrac{3f_0}{2\pi}\dfrac{xz}{r^{5}} + O(h^2) \ .  \label{Imp_pressure}
\end{eqnarray}
A knowledge of $p$ makes it possible to calculate the resulting free surface displacement.
Below, we establish the free surface displacement in response to a given surface pressure field. 
The velocity potential $\phi(\boldsymbol r ,t)$ is determined by solving the Laplace equation $\Delta \phi =0$ together with the boundary conditions:
\begin{subeqnarray}
\boldsymbol \nabla \left[ {\partial_t^{2}\phi}+g{\partial_z\phi}-\gamma {\partial_z}\Delta_{x,y}\phi +\frac{1}{\rho}{\partial_t P}  \right]_{z=0}&=&\boldsymbol 0  \slabel{BCsurface}\\
 \partial_z \phi \big|_{z\rightarrow -\infty} &=& 0\ ,
\end{subeqnarray}
where $g$ denotes the acceleration of gravity, $\gamma$ the surface tension, and $P(\boldsymbol r ,t)$ the imposed pressure field. The general solution is (\cite{Raphael1996}):
 \begin{eqnarray}
 \phi(\boldsymbol{r},t)&=&\int \frac{\text d^{2}\boldsymbol{k}}{(2\pi)^2}\, A(\boldsymbol{k},t)\,e^{i (u x+vy)}e^{kz} \ , \label{Gensol}
\end{eqnarray}
 where $\boldsymbol k = (u,v)$. Projecting Eq.~\eqref{BCsurface} along the vertical axis $\boldsymbol e_z$ and incorporating Eq.~\eqref{Gensol} leads to:
  \begin{eqnarray}
k \left[{\partial_t^{2}A}+\omega(k)^{2}A\right]&=&-\frac{1}{\rho} {\partial_t \mathcal F\left[{\partial_z P\big|_{z=0}}\right]} \ ,
\label{eqnA}
\end{eqnarray}
where $\omega^2=\gamma k^3/\rho + gk$ denotes the dispersion relation for capillary-gravity waves, and  $\mathcal F$ denotes the spatial two-dimensional Fourier transform operator (see Appendix A). The kinematic boundary condition $\partial_t\zeta = \partial_z\phi|_{z=0}$ can be written in Fourier space as:
  \begin{eqnarray}
  \partial_t \hat \zeta &=& kA \ , \label{BCkin}
  \end{eqnarray}
 where $\hat \zeta (\boldsymbol k,t)= \mathcal F [\zeta]$. Incorporating Eq.~\eqref{BCkin} into Eq.~\eqref{eqnA} and integrating over time yields:
  \begin{eqnarray}
{\partial_t^{2}\hat \zeta}+\omega(k)^{2}\hat \zeta&=&-\frac{1}{\rho} { \mathcal F\left[{\partial_z P\big|_{z=0}}\right]} \ .
\label{eqnZetahat}
\end{eqnarray}
Equation~\eqref{eqnZetahat} is of prime importance in that it can be used to calculate the surface displacement $\zeta(\boldsymbol x,t) =\mathcal F^{-1}[\hat \zeta]$ for any given pressure distribution $P(\boldsymbol r,t)$ and for any given set of initial conditions.
The general solution can be written as the sum of the solution to the homogeneous equation and a particular solution to the non-homogeneous equation, namely $\hat{\zeta}(\boldsymbol k,t) =  \hat{\zeta_{\text{h}}}(\boldsymbol k,t) + \hat{\zeta_{\text{p}}}(\boldsymbol k,t)$ where:
 \begin{subeqnarray}
 \hat{\zeta}_{\text{h}}(\boldsymbol{k},t)&=& \hat{\zeta}(\boldsymbol{k},0)\cos (\omega(k)t)+\partial_{t}\hat{\zeta}(\boldsymbol{k},0)\dfrac{\sin(\omega(k) t)}{\omega(k)}  \\
  \hat{\zeta}_{\text{p}}(\boldsymbol{k},t)&=&-\frac1\rho\int_{0}^{t}\text{d}\tau\, \frac{\sin (\omega(k)(t-\tau))}{\omega(k)} \,{\mathcal F\left[{\partial_z P}\big|_{z=0}\right]} \ , \slabel{partsol}
\end{subeqnarray}
and the particular solution is determined by the variation of constants method (\cite{Lucas}).  The homogeneous solution is the well-known \textit{Cauchy-Poisson} solution, also known as the free-wave solution (\cite{lamb1993hydrodynamics}). We assume here that the free surface is initially considered to be at rest, with $\hat{\zeta}(\boldsymbol{k},0)=0$ and $\partial_{t}\hat{\zeta}(\boldsymbol{k},0)=0$, such that $\hat{\zeta}(\boldsymbol k,t) = \hat{\zeta_{\text{p}}}(\boldsymbol k,t)$. The impulsive pressure distribution takes the form $P(\boldsymbol{r},t)=p(\boldsymbol{r})\delta(t)$, where $p(\boldsymbol{r})$ is given by Eq.~\eqref{Imp_pressure}. Incorporating $P(\boldsymbol r,t)$ into Eq.~\eqref{partsol} yields:
 \begin{eqnarray}
 \hat{\zeta}(\boldsymbol{k},t)&=&-\frac1\rho \dfrac{\sin(\omega(k)t)}{\omega(k)} \, \mathcal F\left[{\partial_z p}\big|_{z=0}\right]  \ , \label{solpartzetahat}
\end{eqnarray} 
where, at first order, in $h$:
 \begin{eqnarray}
\mathcal F\left[{\partial_z p}\big|_{z=0}\right]&=&-\mathcal F \left[\dfrac{3f_0}{2\pi}\dfrac{hx}{\left(x^2+y^2+h^2\right)^{5/2}}\right]  \ .\label{pressure}
\end{eqnarray} 
 Fourier transform in polar coordinates (see Appendix A) yields 
${\mathcal F\left[{\partial_z p}\big|_{z=0}\right] = if_0 k_x\,e^{-hk}}$, such that taking the inverse Fourier transform of Eq.~\eqref{solpartzetahat} and changing to a polar set of integration variables finally yields: 
\begin{eqnarray}
{\zeta_{\text{imp}}(\boldsymbol x,t)=\frac{f_0}{2\pi\rho }\dfrac{x}{\sqrt{x^{2}+y^{2}}}\int_{0}^{\infty}\text{d}k\,  k^{2}e^{-hk} J_{1}\left (k \sqrt{x^{2}+y^{2}}\right )\dfrac{\sin(\omega(k)t)}{\omega(k)} }\ . 
\label{Solzeta}
\end{eqnarray}
Equation \ref{Solzeta} provides the change in the free surface after an impulsion.

\subsection{Finite depth}

As indicated above, the measurements must be performed in shallow water due to the constraints of optical resolution. We therefore extend the theory to the case of shallow water. With the same assumptions as above, we consider an impulsive force field located at $x=y=0$ and depth $z=-h$, where $h>0$ in shallow water of depth $H>h$.
The force field is given by Eqs (\ref{expF}) and (\ref{expf}), and the resulting pressure field is found by solving the Poisson equation (see Eq.~(\ref{Poisson})) but with a different set of boundary conditions: $p(z=0)=0$, $\partial_z p(z=H)=0$ and $\boldsymbol \nabla p|_{r\rightarrow \infty}\rightarrow \boldsymbol 0$. For the sake of readability, the details of the calculations are reported in  Appendix A.  The solution for the pressure field is as follows (see Appendix A):
\begin{equation}
{	p_h=-\frac{f_0}{\pi H}\cos(\theta)\sum_{n=0}^{\infty}\Lambda_n\sin\left(\Lambda_n z\right)\sin\left(\Lambda_n h\right)K_{1}\left(r\Lambda_n \right) \ ,}
\end{equation}
where $\Lambda_n = (2n+1)\pi/(2H)$. The change in the free surface, calculated as described above, is given by (see Appendix A):
\begin{equation}
{	\zeta_{\text{imp}}^H(\boldsymbol{x},t)=\frac{f_0}{2\pi\rho}\frac{x}{r}\int_{0}^{+\infty}\text{d}k k^2 \left[{\cosh(kh)-\sinh(kh)\tanh({Hk})} \right] J_1(kr)\frac{\sin(\omega(k)t)}{\omega(k)} }\ ,
	\label{Solzetashallow}
\end{equation}
where $\omega^2= (\gamma k^3/\rho +gk)\tanh(kh)$ is  the dispersion relation for finite-depth capillary gravity waves \cite{lamb1993hydrodynamics}.
In deep water (i.e. $H\to \infty$), we recover Eq.~(\ref{Solzeta}) (see Appendix A).

\section{Towards realistic stroke modelling}

Here, we compare  experimental observations of  surface displacement to the theoretical predictions developed in the previous section. This leads to the replacement of impulsive  forcing by continuous forcing. We then fit the theoretical surface to the experimental surface, with the amplitude of the propulsive force $f_0$ (see Eq.~\eqref{expf}) as output. 

\subsection{Qualitative comparison, from impulsive to continuous forcing}

Figure~\ref{F7} (left column) displays the time-resolved surface topography around an individual of \textit{Gerris paludum}, during the 10 ms of a leg stroke. The central column shows a theoretical estimate of topography according to B\"uhler's method for impulsive forcing. It is clear at a glance that the lack of agreement between theory and experimental results leads to  poor qualitative prediction of the wave pattern.
%
Indeed, the experiments showed the wave to be of an elongated circular shape, reflecting the circular displacement of the leg during the stroke, whereas theory predicted a symmetric shape\footnote{Furthermore, over a period of 1 to 10 ms, the wave produced after the impulse in experimental conditions increased, reaching a maximal value at 8 ms. By contrast, the theoretical prediction of surface elevation given by  Eq.~(\ref{Solzeta})  yields a maximal wave crest at 4 ms (\cite{buhler}).}. In addition, theory predicted many more oscillations than were observed in the experiment. Interestingly, this particular feature can be resolved by accounting properly for finite depth, consistent with the shallow-water conditions in which the experiments were performed. Another small improvement can be obtained by taking into account the length of the strider's leg in contact with the water surface (leg length is $l_{\text{leg}}=3.75$ mm).\\

\begin{figure}
\centering
\includegraphics[scale=0.30]{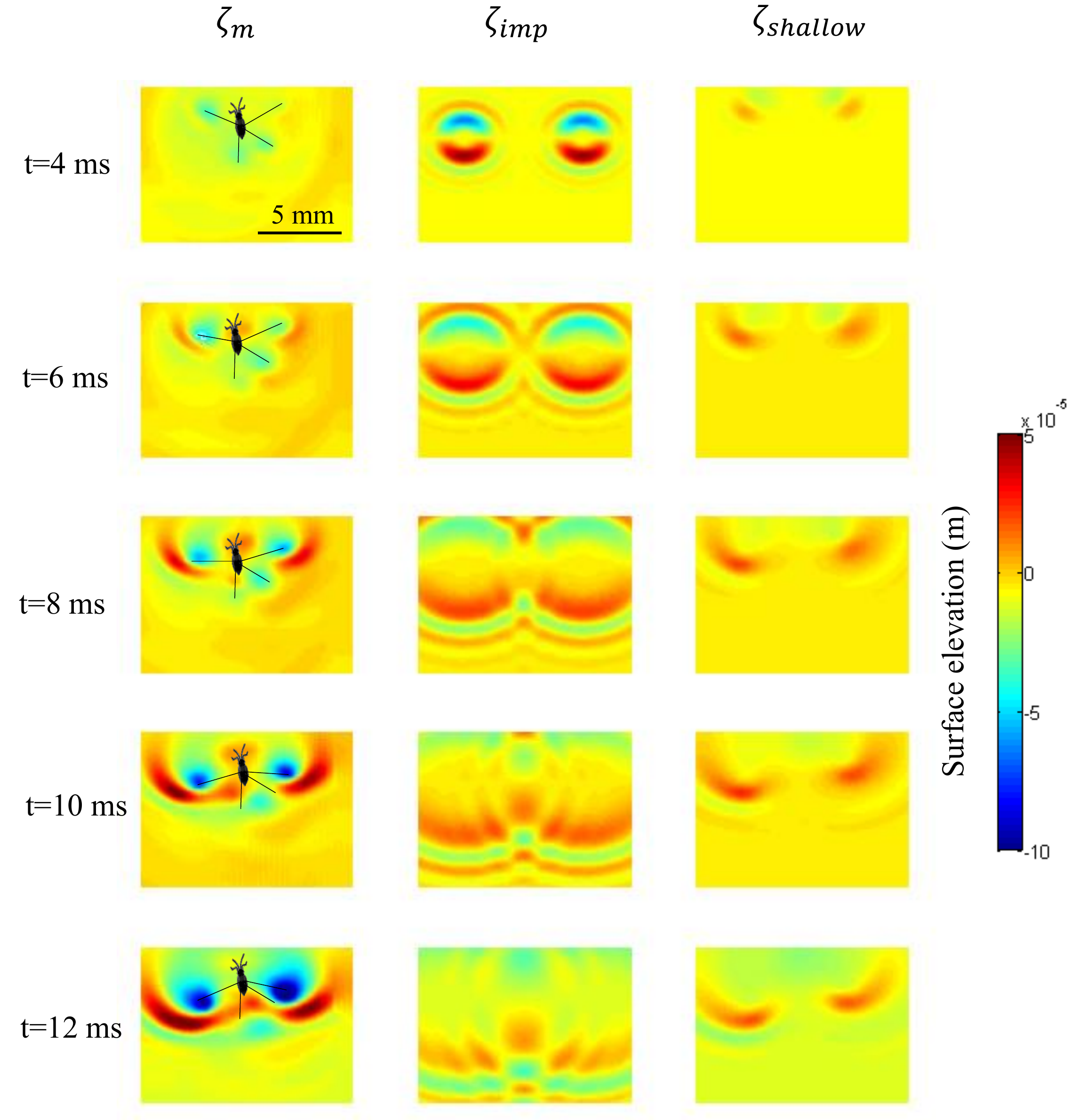}
\caption{Comparison of the experimentally measured surface elevation with theoretical predictions. The 1$^{st}$ column represents the experimental surface elevation generated during the propulsion of a \textit{Gerris paludum} individual ($m_\text{gerris}=0.348$ mg, $V_\text{gerris}=0.38$ m.s$^{-1}$).  The 2$^{nd}$ column represents the theoretical surface elevation produced by a single impulse in the direction of motion, with $\zeta_{\text{imp}}$ as expressed in equation \eqref{Solzeta} (\cite{buhler}). The 3$^{rd}$ column shows the results for  continuous forcing in \textit{shallow water} along the trajectory of the water strider legs (see Eqs.~\eqref{zetacont} and \eqref{zetacontdis}).}
\label{F7}
\end{figure}

However, a real improvement  emerges from the observation that impulsive forcing is a very rough approximation of the real continuous forcing of the legs. During a leg stroke, the leg tip imparting momentum to the interface is not localised in space but instead follows a circular trajectory during the duration of the stroke. Accounting for the temporal extension of the rowing period, consistent with the linearity of the governing equation, we superimposed the impulsive solutions along the trajectory of the leg. Denoting $\boldsymbol x_{\text{leg}}(t)  =  (x_{\text{leg}}(t),y_{\text{leg}}(t))$ for ${t\in\left[0,\tau_{\text{imp}}\right]}$ the trajectory of a leg during the stroke phase, we obtain:
\begin{eqnarray}
\zeta_{\text{cont}}(\boldsymbol x,t)=\int_0^t   \frac{\text d s}{\tau_{\text{imp}}} \,\Theta({s<\tau_{\text{imp}}})\, \zeta^H_{\text{imp}}(\boldsymbol x-\boldsymbol x_{\text{leg}}(s),t-s) \ ,
\label{zetacont}
\end{eqnarray}
where $\Theta$ is the Heaviside step function, and  $\zeta^H_{\text{imp}}$ is given by Eq.~\eqref{Solzetashallow}.
 The time-dependent positions of the tips of the left and right legs were obtained by fitting two parametric equations $\boldsymbol x_{\text{leg}}(t)$=$(x_{\text{leg}}(t),y_{\text{leg}}(t))$.
We calculated  the total surface deformation numerically, by discretising Eq.~(\ref{zetacont}) as follows:
 \begin{equation}
\zeta_\text{cont}(\boldsymbol x,\tau)=\sum_{n=0}^{N}\Theta\left(\tau-n\Delta t\right)\zeta_\text{imp}\left(\boldsymbol x-\boldsymbol x_{\text{leg}}(n\Delta t),\tau-n\Delta t \right)\frac{\Delta t}{\tau_\text{imp}} \ , \label{zetacontdis}
\end{equation}
with $\Delta t$ the discretisation time step, $N={\tau_\text{imp}}/{\Delta t} $ the number of time iteration steps, and $\Theta$ the Heaviside step function. The results are displayed in the right column of Fig.~\ref{F7}. As we can see, the typical elongated circular wave produced by the legs is very well reproduced by the simulation of continuous forcing in shallow water. 
%

\subsection{Quantitative fitting}

As described above, fitting the theoretical surface to the experimental data by minimising the mean square difference makes it possible to obtain the total exerted force  $f_0$ as a fitting parameter.
We obtain\footnote{Note that the units of this force are quite uncommon. Indeed, an impulsive forcing involves delta function. In the definitions provided in Eqs.~\eqref{expF} and \eqref{expf}, the delta functions have the inverse dimension of their arguments. As such, $f_0$ is a force multiplied by a volume multiplied by a time.} ${f_0}  =  1.3 \times 10^{-10} {\text{ kg}}.\text{m}^4 .\text s^{-1}$. 

\section{Discussion}

{\subsection{Modelling of the forcing}
As mentioned above, the modelling of leg forcing through a localised pressure field below a flat surface represents a major simplification of reality. This approach was validated with hindsight, as the agreement between experiments and theoretical predictions (see Fig. 3) was very good, but several aspects of the complex nature of leg kinematics and fluid-structure interaction merit further discussion.\\
}

{ In reality, the surface is initially deformed by the weight of the insect (see Fig.~1a).
 Accounting for this initial deformation of the water surface by the leg would probably slightly improve the agreement between theory and experimental results, particularly over short time periods. The instantaneous almost static deformation of the surface could be incorporated, by monitoring leg depth at given time points in during propulsion. A two-dimensional approach of this kind was proposed by \cite{gao2011numerical}. However, the dynamic deformation of the interface is complex in three dimensions, and it is beyond the scope of this study to reproduce the surface deformation with minimal input in the simplest possible framework, to extract the main physical elements of strider locomotion.}\\
 
 {Another point worth discussing is the direction of the forcing. In this work, we have neglected the vertical component of the forcing, by assuming a purely horizontal impulse. \cite{buhler}  addressed the matter of oblique recoil forces and showed that the ratio of the vertical and horizontal components of the forcing can be estimated through the total vertical and horizontal displacements of the driving legs during the propulsion phase. In our experiments (see Fig. 3), there is a vertical displacement of 100 $\mu$m and a horizontal displacement of 5 mm, yielding an oblique forcing angle of about 1 degree, which can be considered negligible\footnote{ {Note that this is not the case for the basilisk lizard, in which vertical forcing must be accounted for (see \cite{glasheen1996hydrodynamic}).}}.}

\subsection{Limitations of our wave recording technique}

There is a trade-off between spatial resolution and the dimensions of the field of view, which we address here. Leg strokes lead to the animal gaining momentum. A proportion of this momentum is transferred to the interface and eventually spreads outside the  field of view for the measurement, after a time determined by  wave velocity and the dimensions of the measurement field. We estimated that the waves produced by the two median legs travelled at $\approx 30$ cm/s. Figure \ref{F7} shows the local elevation of the air-water interface $\zeta(x,y)$ on a measurement field extending over an area of $18\times13$ mm. The  waves are produced in the middle of the measurement field, which they leave  $\approx 30$ ms after the beginning of the stroke.  This time period is longer than the duration of the impulse $\tau_\text{imp}$, and field size is, therefore, satisfactory.  For measurements on 3rd instars , the measurement field was set to $43 \times 31$ mm. With a mean wave speed of 50 cm/s, the waves  take 50 ms to leave the measurement field. Increasing the size of the field of view would make it possible to follow the wave for longer after its production, but at the expense of spatial resolution.

\subsection{Comparison with previous studies}

Table \ref{tab_comp} presents the main characteristics of our measurements, together with those of two previous studies (\cite{hu2003hydrodynamics,gao2011numerical}). Insect momentum per unit mass of fluid acquired during propulsion (as defined by B\"uhler) was estimated through two different approaches: a kinematic analysis and a fitting of the numerical solution of surface deformation. \cite{hu2003hydrodynamics} chose to estimate the momentum of the wave with slowly varying wave train theory.  They obtained a larger wave crest than reported here. The use of relatively large insects (body mass 10 mg) and a higher velocity (1 m/s) also resulted in a relatively large body momentum. However, the use of an essentially steady wave theory for estimating the momentum of unsteady waves was called into question by \cite{buhler}.  \cite{gao2011numerical} subsequently revisited the question, focusing their model on the interaction of an infinite cylinder with a 2D air-water interface, and using a numerical approach. They did not model the motion of an entire body, thereby avoiding complications relating to body velocity, mass, acceleration or momentum. A comparison with our work is, therefore, not straightforward. The numerical values of forces, expressed in N/m due to the 2D nature of their work, must be corrected by the leg/interface contact length, for comparison with our estimates. The two studies also differ in terms of leg diameter and stroke depth, accounting for the higher Reynolds number and wave crest elevations than reported here.\\
%


\begin{table*}
    \centering
\resizebox{1\textwidth}{!}{  \includegraphics{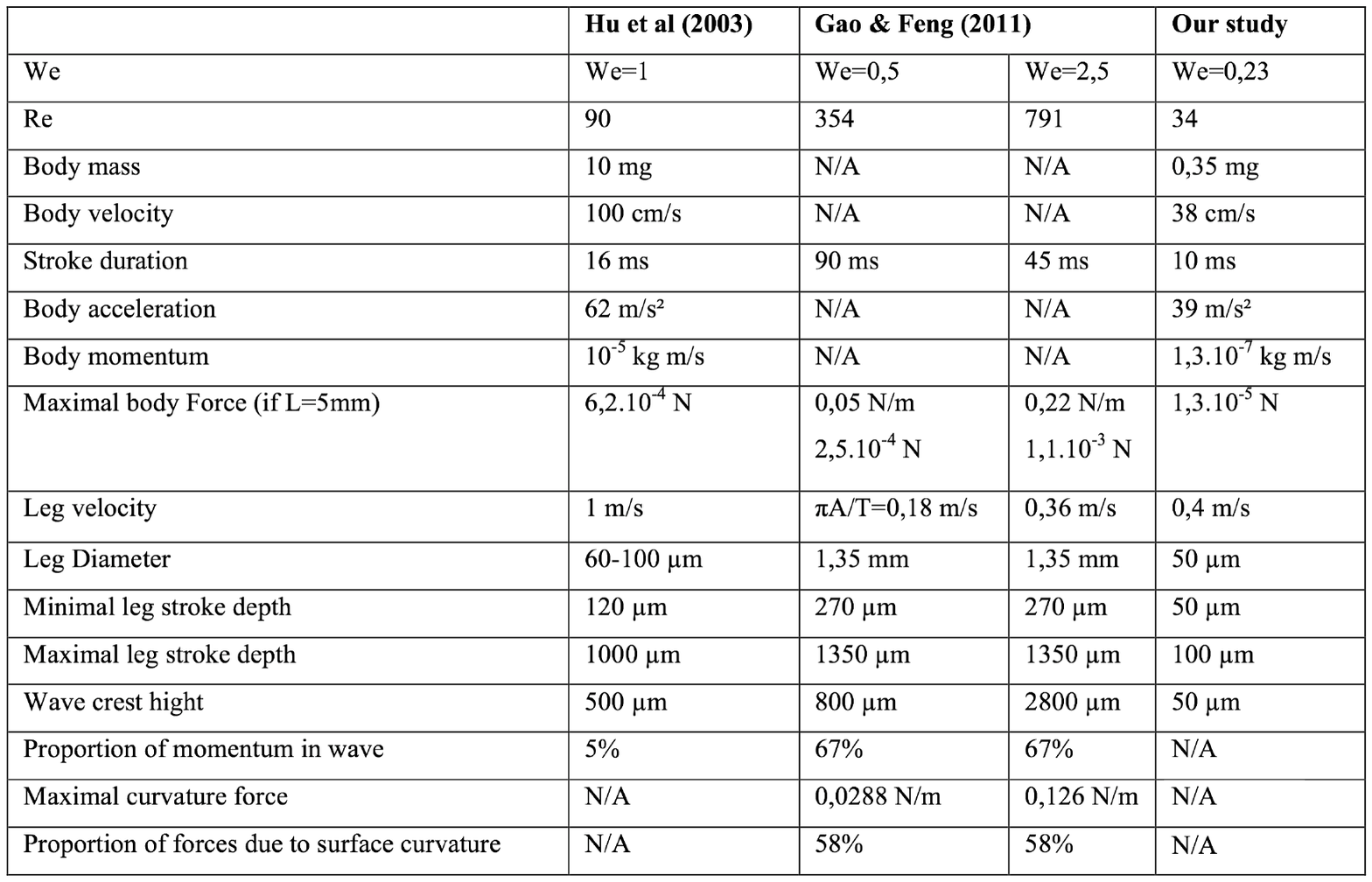}}\medskip
\caption{Table comparing the main characteristics of our measurements with those of two previous studies (\cite{hu2003hydrodynamics,gao2011numerical}). }
\label{tab_comp}     
\end{table*}

\subsection{Correction of B\"uhlerÕs theory for finite depth}

We believe that the extension of B\"uhler's theory to finite depth is also of potential interest for other situations, such as momentum transfer in thin films.
In their recent work on the angled impact of a drop impinging on a flowing soap film, \cite{basu2017angled} pointed out the remarkable similarities between this situation and the impulsive interaction observed in water strider locomotion. In their experimental setup, a drop hitting the surface of a soap film  sheds a vortex dipole and generates capillary waves. The authors were able to quantify the expected loss of vortex dipole momentum derived from B\"uhler's theory, from the measured pre- and post-impact drop velocities and the experimental vortex momentum,  calculated from particle image velocimetry data. The estimates were very different, with B\"uhler's theory underestimating the momentum in the vortices by a factor of six. In addition to the explanations provided by the authors, combining a distinct trampoline effect and the bulk elastic deformation of the soap film leading to additional restitutive forces, it might also be of interest to compare the results obtained with the finite depth theory presented here. 


\section{Concluding remarks}

We present here experiments and theoretical considerations concerning the surface displacement generated by the leg strokes of water striders. Using the synthetic Schlieren method, we performed dynamic measurements of  the topography of the air-water interface during the water strider's leg strokes, with unprecedented accuracy. We extended B\"uhler's study to  finite depth, consistent with experimental conditions. We found that it was necessary to account for continuous forcing along the trajectory of the leg to reproduce the experimental wave patterns. This suggests that in the real conditions prevailing in the natural habitat of \textit{Gerris} (including infinite depth: $H\gg \lambda$), B\"uhler's approach is highly satisfactory, provided that leg length is accounted for and the impulsive solution is superimposed along the entire trajectory of the leg. 
\\

This study constitutes a step towards a fine quantitative understanding of the mechanism of locomotion in water striders. It revealed substantial variability in the speed, weight and leg kinematics of individuals, together with a large impact of the developmental stage (instar) of the animal, as shown in static conditions. These  aspects of \textit{Gerris} locomotion affect hydrodynamic aspects of locomotion, including the partitioning of momentum between waves and vortices. In our view, future studies should rigorously test B\"uhler's prediction of momentum partition, one of the key questions relating to propulsion at the air-water interface.  B\"{u}hler concluded that one third of the momentum was distributed in waves, and the other two thirds in  vortices. By contrast, \cite{gao2011numerical}, reported almost the opposite situation, based purely on computation, and noted that the relative importance  of waves and vortices depended on the size of the animal. Size-dependent force partitioning was also observed by \cite{hu2010hydrodynamics}, who showed that small instar striders brushed the surface and were unable to deform the meniscus: the viscous forces propel them forward. {To our knowledge, there is no conclusive experimental evidence definitively settling this question of partitioning. } In terms of biomechanical understanding, only a full energy balance, including metabolic expenditure, would provide a conclusive demonstration settling the debate relating to the partitioning of momentum between waves and vortices, and its dynamics over time and between instars. Future studies should investigate the energy lost in the fluid under the surface, to provide a correct estimate of the complete energy balance. Recent advances in tomo-PIV, including improvements in the ability to follow Lagrangian particle movement, should make the quantitative study of the small vortices created feasible in 3D, despite the difficulty of imaging immediately beneath a moving air-water interface (\cite{sveen2004quantitative}), particularly under a small rapidly moving object.\\

We thank Fr\'ed\'eric Moisy for lending the plexiglas model, and Ulysse Mizrahi, Romain Paserot and Alexandre Darmon for fruitful discussions.


\newpage

\section*{Appendix A. Pressure field and oriented impulse solutions}

\subsection*{A.1. Infinite depth}

Recalling the results of \cite{buhler}, we compute here the response of the free surface to the impulsive pressure field in water of infinite depth.
Taking the pressure field given by Eq.~(\ref{pressure}) and computing the Fourier transform with polar coordinates ($r,\theta$), we eventually obtain:\footnote{{Here we have used $\mathbf k\cdot\mathbf r =kr\cos(\theta-\phi)$, with $\theta$ the polar coordinates in real space and $\phi$ the polar coordinates in Fourier space.} }
\begin{eqnarray}
\mathcal{F}\left[\dfrac{\partial p_{\text{imp}}}{\partial z}\big|_{z=0}\right]&=& -\frac{3f_0 h}{2\pi}\iint \text{d}x\text{d}y \ \frac{x}{(x^2+y^2+h^2)^{5/2}}e^{-i(k_xx+k_yy)}\nonumber \\
&=& -\frac{3f_0 h}{2\pi}\iint \text{d}r\text{d}\theta \ \frac{r^2\cos(\theta)}{(r^2+h^2)^{5/2}}e^{-ikr\cos(\theta-\phi)}\nonumber \\
&=&-\frac{3f_0 h}{2\pi}\int_0^{\infty} \text{d}r \ \frac{r^2}{(r^2+h^2)^{5/2}}\int_{-\pi}^{\pi}\cos(\theta)e^{-ikr\cos(\theta-\phi)} \ .
\end{eqnarray}
After integrating over theta, we obtain:
\begin{eqnarray}
	\int_{-\pi}^{\pi}\frac{\text{d}\theta}{2\pi}\cos(\theta)e^{-ikr\cos(\theta-\phi)} &=&-i\cos(\phi)J_{1}(kr) \ .
\end{eqnarray}
Thus 
\begin{eqnarray}
\mathcal{F}\left[\dfrac{\partial p_{\text{imp}}}{\partial z}\big|_{z=0}\right]&=&3i f_0 h{ \frac{k_x}{k}}\int_{0}^{+\infty}\text{d}r\, \dfrac{r^{2}}{(r^{2}+h^{2})^{\frac{5}{2}}}J_{1}(kr)  \nonumber \\
&=&i f_0 k_x \exp(-hk) \ .
\label{TFIDeep}
\end{eqnarray}
\noindent
From this distribution, we can calculate the change in the free surface after an oriented impulsion:
\begin{eqnarray}
\zeta_{\text{imp}}(\boldsymbol x,t)= -i \dfrac{f_0}{(2\pi)^{2}\rho}\int_{0}^{+\infty}\text{d}k_x\text{d}k_y  k_x \exp(-hk)\dfrac{\sin(\omega(k)t)}{\omega(k)}\exp(i(k_xx+k_yy)) \ .
\end{eqnarray}
\noindent
After  polar transformation in Fourier space we obtain an expression for the change in the free surface after an impulsive strike (Eq. \ref{Solzeta}):
\begin{eqnarray}
\zeta_{\text{imp}}(\boldsymbol x,t)= \frac{f_0}{2\pi\rho }\dfrac{x}{\sqrt{x^{2}+y^{2}}}\int_{0}^{\infty}\text{d}k\,  k^{2}e^{-hk} J_{1}\left (k \sqrt{x^{2}+y^{2}}\right )\dfrac{\sin(\omega(k)t)}{\omega(k)} \ .  \label{buhlerinfd}
\end{eqnarray}

\subsection*{A.2. Finite depth}
In the case of finite depth, we must first compute the pressure field resulting from an impulsive force in water of finite depth and the resulting surface response.
Let $G$ be the Green's function of the problem given by Eq~(\ref{Poisson})  subject to $p(z=0)=0$, $\partial_z p(z=-H)=0$ and $\nabla p \rightarrow 0$ as $\sqrt{x^2+y^2}\rightarrow\infty$.
The function $G$  solves:
$	\Delta G=\delta(\boldsymbol{x}-\boldsymbol {x}')$,
with the same boundary conditions, leading to:
	${p= -{f_0}\partial_x G(\boldsymbol {x},0,0,-h)}$.
Using the Eigen decomposition of the Laplacian with these boundaries:
\begin{equation}
	\Psi_n(k_x,k_y)=\frac{\sqrt{2/H}}{2\pi}e^{i\boldsymbol {k}\cdot\boldsymbol {r}}\sin\left(\Lambda_n z\right) \ ,
\end{equation}
where  $\Lambda_n={(2n+1)\pi}/(2H)$, together with the spectral formula of  Green's function, we obtain:
\begin{eqnarray}
	G(\boldsymbol {r},\boldsymbol {r'})&=&\frac{1}{2\pi^2 H}\sum_{n=0}^{\infty}\sin\left(\Lambda_n z\right)\sin\left(\Lambda_n z'\right)\iint dk_xdk_y \frac{e^{i\boldsymbol {k}\cdot\boldsymbol {(r-r')}}}{k^2+\Lambda_n^2} \nonumber \\
&=&\frac{1}{\pi H}\sum_{n=0}^{\infty}\sin\left(\Lambda_n z\right)\sin\left(\Lambda_n z'\right)K_0\left(\vert r-r'\vert\Lambda_n\right) \ .
\end{eqnarray}
The horizontal impulsive pressure field is thus:
\begin{equation}
	{p_h=-\frac{f_0}{\pi H}\cos(\theta)\sum_{n=0}^{\infty}\Lambda_n\sin\left(\Lambda_n z\right)\sin\left(\Lambda_n h\right)K_{1}\left(r\Lambda_n \right) \ .}
\end{equation}
To calculate the deformation of the free surface, we then need to compute the Fourier transform of $\partial_z{p_h}(z=0)$. Computing the derivative with respect to $z$ yields:
\begin{equation}
	\frac{\text{d}p_h}{\text{d}z}\vert_{z=0}=-\frac{f_0}{2\pi H}\cos(\theta)\sum_{n=0}^{\infty}\Lambda_n^2\sin\left(\Lambda_n h\right)K_{1}\left(r\Lambda_n \right) \ ,
\end{equation}
with $r=\sqrt{x^2+y^2}$.
We set $\tilde f(k_x,k_y)$ the Fourier transform of  $\dfrac{\text{d}p_h}{\text{d}z}(z=0)$:
\begin{eqnarray}
	\tilde f(k_x,k_y)&=&\iint \text{d}x\text{d}y \text{ } e^{-i(k_x x+k_y y)}\dfrac{\text{d}p_h}{\text{d}z}\vert_{z=0}\nonumber \\
&=&-\frac{f_0}{\pi H}\sum_{n\geq0}\Lambda_n^2\sin\left(\Lambda_n h\right)  \int_{0}^{+\infty}\text{d}r \ rK_{1}\left(r\Lambda_n \right) \int_{0}^{2\pi} \text{d}\theta\cos(\theta)e^{-ikr\cos(\theta-\phi)}\nonumber \\
	&=&\frac{2if_0}{H}\sum_{n\geq0}\Lambda_n  ^2\sin\left(\Lambda_n  h\right) \cos(\phi)\int_{0}^{+\infty} K_{1}\left(r\Lambda_n  \right)J_1(kr)r\text{d}r \nonumber\\
&=&\frac{2if_0}{H}k_x\sum_{n\geq0} \sin\left(\Lambda_n  h\right)\frac{\Lambda_n}{k^2+\Lambda_n ^2}\ .
\end{eqnarray}
The sum above can be calculated as follows:
\begin{equation}
	\sum_{n=1}^{\infty}\frac{n}{y^2+n^2}\sin(nx)=\frac{\pi}{2}\frac{\sinh(y(\pi-x))}{\sinh(\pi y)} \ .
\end{equation}
which yields:
\begin{eqnarray}
	\tilde f(k_x,k_y)&=&if_0k_x\left(\cosh(kh)-\sinh(kh)\tanh({kH})\right) \ .
\end{eqnarray}
The case of an impulsion in infinitely deep water is a limiting case of this expression:
\begin{equation}
	\lim_{H\to+\infty}	\tilde f(k_x,k_y)=if_0k_xe^{-kh} \ ,
\end{equation}
which is exactly the Eq.~\ref{TFIDeep}.
The displacement of the free surface is thus:
\begin{eqnarray}
\zeta_\text{imp}^{H}(\boldsymbol x,t)= \frac{f_0}{2\pi\rho }\frac{x}{r}\int_{0}^{\infty}\text{d}k\,  k^{2}\left(\cosh(kh)-\sinh(kh)\tanh(kH)\right) J_{1}\left (k r\right )\dfrac{\sin(\omega(k)t)}{\omega(k)} \ .  \label{buhlerinfd}
\end{eqnarray}
where $\omega^2=({k+k^3})\tanh(k{H}/{l_c})$ and $l_c $ is the capillary length. Note that we can easily recover B\"uhler's result in deep water by letting ${H\to+\infty}$.

\newpage

\begin{table*}
\medskip

\section*{Appendix B. Experimental sample table}

    \centering
\resizebox{0.9\textwidth}{!}{  \includegraphics{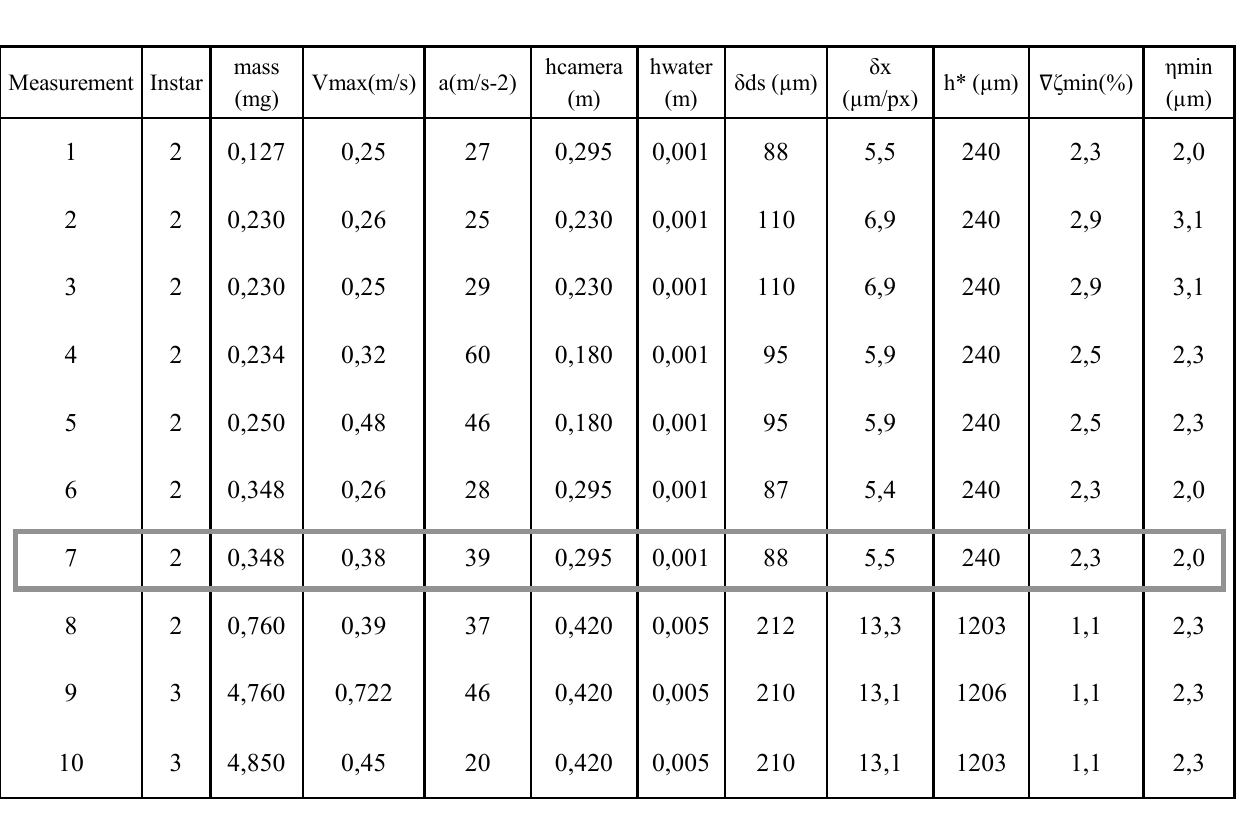}} \medskip
\caption{Table of the physical, kinetic and geometric parameters used to reconstruct the local surface topography. We report  the mass (mg) of each individual, the maximal velocity $V_\text{max}$ (m/s) reached by the insect after its stroke, mean acceleration $a$ (m/$\text{s}^{2}$) during the stroke,  the distance $h_\text{camera}$ (m) between the camera and the water tank,  the water surface elevation at rest $h_\text{water}$ (m),  the distance $\Delta \text{d}s$ ($\mu\text{m}$) between each estimate of elevation,  spatial resolution $\Delta x$ ($\mu\text{m}$/pixels),  the distance $h^*$($\mu\text{m}$) used to estimate the minimum measurable surface gradient $\nabla\zeta_{\text{min}}(\%)$, and $\eta_{min}$ the resulting resolution of the technique for elevation . The gray square indicates the experiment used throughout this manuscript (see Figs.~3 and 7).
}
\label{tab}     
\end{table*}

\clearpage

\bibliographystyle{jfm}
\bibliography{biblio}

\end{document}